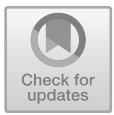

# An End-to-End Structure with Novel Position Mechanism and Improved EMD for Stock Forecasting

Chufeng Li and Jianyong Chen[✉]

School of Computer Science and Software Engineering, Shenzhen University, Shenzhen, China
jychen@szu.edu.cn

**Abstract.** As a branch of time series forecasting, stock movement forecasting is one of the challenging problems for investors and researchers. Since Transformer was introduced to analyze financial data, many researchers have dedicated themselves to forecasting stock movement using Transformer or attention mechanisms. However, existing research mostly focuses on individual stock information but ignores stock market information and high noise in stock data. In this paper, we propose a novel method using the attention mechanism in which both stock market information and individual stock information are considered. Meanwhile, we propose a novel EMD-based algorithm for reducing short-term noise in stock data. Two randomly selected exchange-traded funds (ETFs) spanning over ten years from US stock markets are used to demonstrate the superior performance of the proposed attention-based method. The experimental analysis demonstrates that the proposed attention-based method significantly outperforms other state-of-the-art baselines. Code is available at https://github.com/DurandalLee/ACEFormer.

## 1 Introduction

Stock trend prediction is an important research hotspot in the field of financial quantification. Currently, many denoising algorithms and deep learning are applied to predict stock trends [1]. The Fractal Market Hypothesis [2] points out that stock prices are nonlinear, highly volatile, and noisy, and the dissemination of market information is not uniform. What's more, if future stock trends can be accurately predicted, investors can buy (or sell) before the price rises (or falls) to maximize profits. Therefore, accurate prediction of stock trends is a challenging and profitable task [3,4].

As early as the end of the last century, Ref. [5] exploited time delay, recurrent, and probabilistic neural networks (TDNN, RNN, and PNN, respectively) to forecast stock trends, and showed that all the networks are feasible. With the rapid development of deep learning, many deep learning methods, especially RNN and Long Short-Term Memory (LSTM) [6], have been widely used in the field of financial quantification. Nabipour et al. [7] proved that RNN and LSTM outperform nine machine learning models in predicting the trends of stock data. With the proposal of the attention mechanism [8], such as Transformer [9] which is based on the attention mechanism and has achieved unprecedented results in the field of natural language processing, the focus of time series research has also shifted to the attention mechanism. Zhang et al. [10] proved that, in the field of stock forecast, LSTM combined with Attention and Transformer only had subtle differences, but both better than LSTM. Wang et al. [11] combine graph attention network with LSTM in forecasting stock and get a better result. Ji et al. [12] build a stock price prediction model based on attention-based LSTM (ALSTM) network. But, unlike time series data such as traffic, due to the trading rules of the stock market, the time interval of stock data is not regular. The self-attention mechanism has the ability to focus on the overall relevance of the data, but it is not only weak to capture short-term and long-term features in multi-dimensional time series data [13] but also weak to extract and retain positional information [14]. However, positional information is very important for time series.

Quantitative trading seeks to find long-term trends in stock volatility. However, short-term high-frequency trading can conceal the actual trend of the stock. This means that short-term high-frequency trading is a kind of noise that prevents the right judgment of long-term trends. Highly volatile stock data greatly affects the effectiveness of deep learning models. In the current stock market, there are many indicators used to smooth stock data and eliminate noise, such as moving averages. However, these indicators are usually based on a large amount of historical stock data. They are lagging indicators [15] and cannot timely reflect the actual fluctuations of the long-term trend. In the field of signal analysis, algorithms such as Fourier Transform (FT), Wavelet Transform (WT), and Empirical Mode Decomposition (EMD) can effectively eliminate signal noise and avoid lag. Compared with FT [16] and WT [17], EMD has been proven to be more suitable for time series analysis [16,17]. It is a completely data-driven adaptive method that can better handle non-linear high-noise data and eliminate data noise. However, EMD also has disadvantages [18] such as endpoint effects and modal aliasing.

Since short-term high-frequency trading has great impact on the long-term trend of stocks, removing short-term trading noise can effectively increase the likelihood of the model finding the correct rules for long-term trends. To solve this problem, we introduced a denoising algorithm called Alias Complete Ensemble Empirical Mode Decomposition with Adaptive Noise (ACEEMD). The noise is eliminated by removing the first intrinsic mode function (IMF) [19]. ACEEMD not only solves the endpoint effect problem but also avoids over-denoising and effectively keeps key turning points in stock trends. In this paper, we propose a stock trend prediction solution, **ACE**EMD Attention **Former** (ACEFormer). It mainly consists of ACEEMD, time-aware mechanism, and attention mecha-nism. The time-aware mechanism can overcome the weak ability of the attention mechanism to extract positional information and the irregularity of stock data intervals. The main contributions of this paper are summarized as follows:

- We propose a stock trend prediction solution called ACEFormer. It consists of a pretreatment module, a distillation module, an attention module, and a fully connected module.
- We propose a noise reduction algorithm called ACEEMD. It is an improvement on EMD which can not only address the endpoint effect but also preserve the critical turning points in the stock data.
- We propose a time-aware mechanism that can extract temporal features and enhance the temporal information of input data.
- We conduct extensive experiments on two public benchmarks, NASDAQ and SPY. The proposed solution significantly outperforms several state-of-the-art baselines such as Informer [20], TimesNet [21], DLinear [22], and Non-stationary Transformer [23].

## 2  Methodology

In this section, we first present the proposed ACEFormer. Next, we introduce the noise reduction algorithm ACEEMD. Finally, the time-aware mechanism is designed.

### 2.1  ACEFormer

The architecture of our proposed model is shown in Fig. 1 which includes pretreatment module, distillation module, attention module and fully connected module.

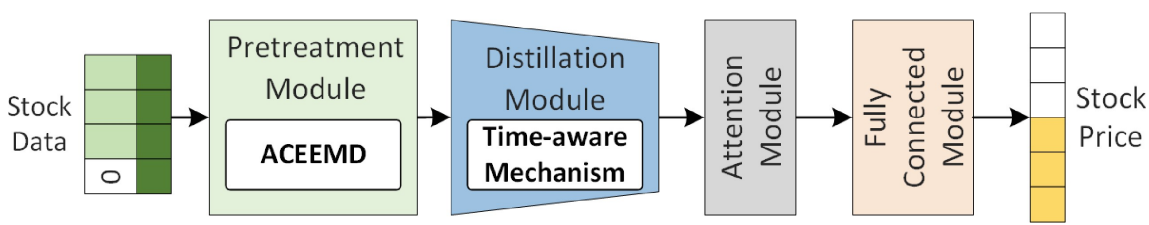

**Fig. 1.** The architecture of ACEFormer.

The pretreatment module preprocesses the input data, which is conducive to the model for better extracting the trend rules of stock data. Among them, our proposed ACEEMD algorithm is also added to the pretreatment module which is shown in Fig. 2.

Let $S = \{s_1, s_2, ..., s_n\}$ represent the stock data input for model training, where $s_i$ includes the price and trading volume of the stock itself and two overall stock market indices on the $i$-th day. Since the data to be predicted is unknown, it is replaced with all zeros. Let $[0:p]$ represent a sequence of $p$ consecutive zeros, where $p$ is the number of days to be predicted. The input data for the model is $D = S||[0:p]$. Let $f$ and $g$ denote functions, $*$ denotes the convolution operation, and $PE(\cdot)$ denotes position encoding. We define the output of pretreatment module denoted by D which is given by:

$$X_{pre} = ACEEMD((f * g)(D)) + PE(D) \tag{1}$$

The distillation module extracts the main features using the probability self-attention mechanism, and reduces the dimension of the feature data using convolution and pooling. In addition, the time-aware mechanism in it is used to extract position features to increase the global position weight.

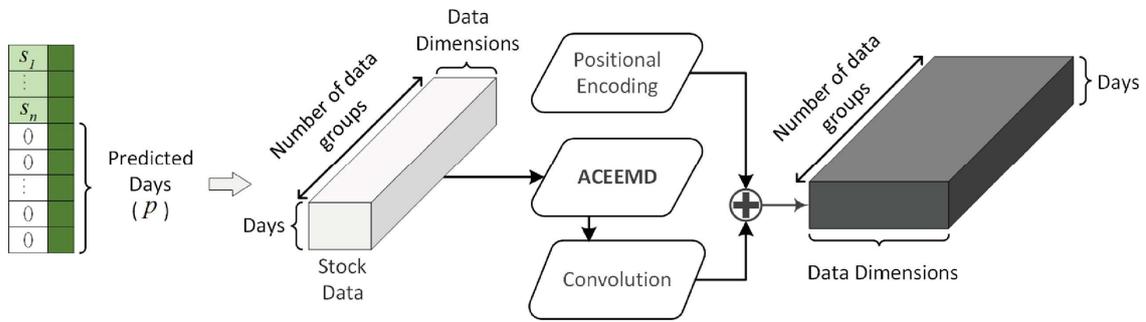

**Fig. 2.** The pretreatment module of ACEFormer.

The distillation module, as shown in Fig. 3, includes the probability attention [20], the convolution, the max pooling, and the time-aware mechanism which is described in detail in the Sect. 3.3. The output features of probability attention contain features of different levels of importance, so the convolution and pooling allow the main features to be retained and the dimensionality of the data to be reduced. It can reduce the number of parameters afterward.

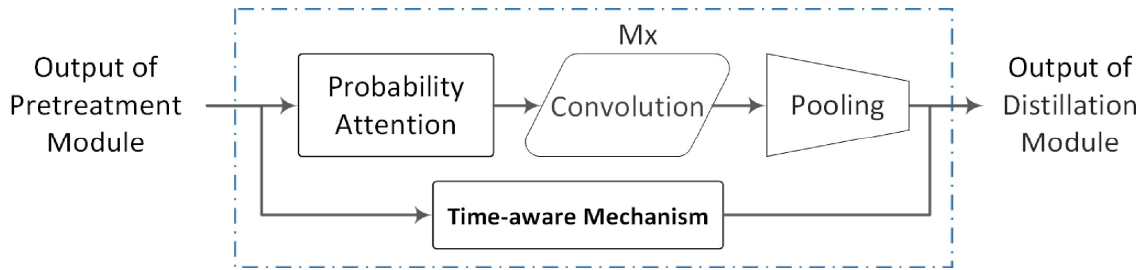

**Fig. 3.** The distillation module of ACEFormer.

The attention module is used to further extract the main features. It can focus on the feature data from the distillation module and extract more critical features. The fully connected module is a linear regression which produces the final predicted values.

Because dimension expansion can generate redundant features, the use of probability attention can increase the weight of effective features based on dispersion. Meanwhile, the convolution and the pooling can eliminate redundant features. In addition, the position mechanism can retain valid features which may be unintentionally eliminated. Because the whole process progressively extracts important features and reduces dimensions of stock data, the self-attention only gets features from the distillation. In the case, it can focus on the compressed data and extract more critical features. Meanwhile, it can effectively eliminate out irrelevant features.

### 2.2 ACEEMD

The ACEEMD can improve the fitting of the original curve by mitigating the endpoint effect and preserving outliers in stock data, which can have significant impacts on trading.

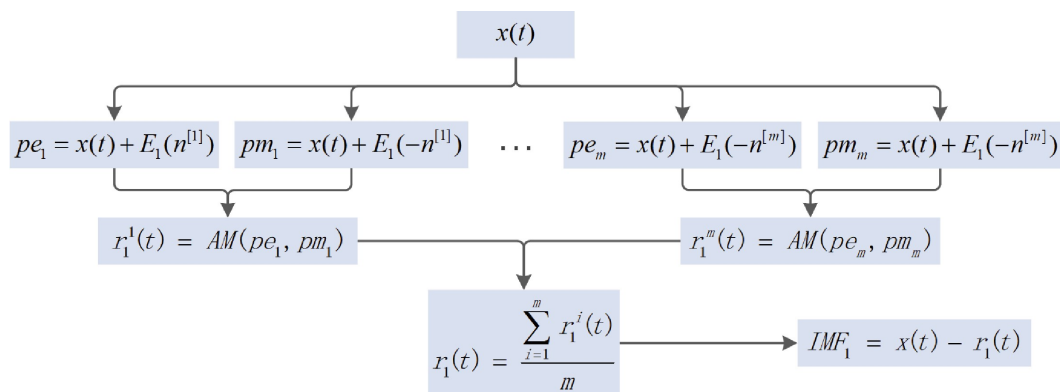

**Fig. 4.** The flowchart of the ACEEMD algorithm architecture.

Figure 4 shows the ACEEMD algorithm. $x(t)$ refers to the input data, i.e. the stock data. $n^{[i]}$ represents the $i$-th Gaussian noise, where the total number of noises $m$ is an adjustable parameter and the default value is 5. $E(\cdot)$ denotes the first-order IMF component of the signal in parentheses. $pe_i$ and $pm_i$ both represent the result of the input data and a Gaussian noise, but the difference between them is that the Gaussian noise they add is opposite in sign to each other. The generating function $AM(pe_i, pm_i)$ is used to denoise the data and is also the core of ACEEMD. $IMF_1$ represents the first-order IMF component of ACEEMD, which is the eliminable noise component in the input data. $r_1^i(t)$ represents the denoised data of the input data with the $i$-th group of added Gaussian noise, and $r_1(t)$ represents the denoised data obtained by processing the input data with ACEEMD.

ACEEMD algorithm has two improvements. First, to avoid the endpoint effect, the input data points for cubic interpolation sampling are constructed using the endpoints and extreme points of the input data. Second, the middle point of a sequence is defined as the data corresponding to the midpoint of the abscissa between the peaks and troughs. The paired data sets with opposite Gaussian noise are $pe_i$ and $pm_i$ shown in Fig. 4. To further preserve the short-term stock trend, the input data points for cubic interpolation sampling of $pm_i$ include not only the extreme points and the endpoints, but also the middle points.

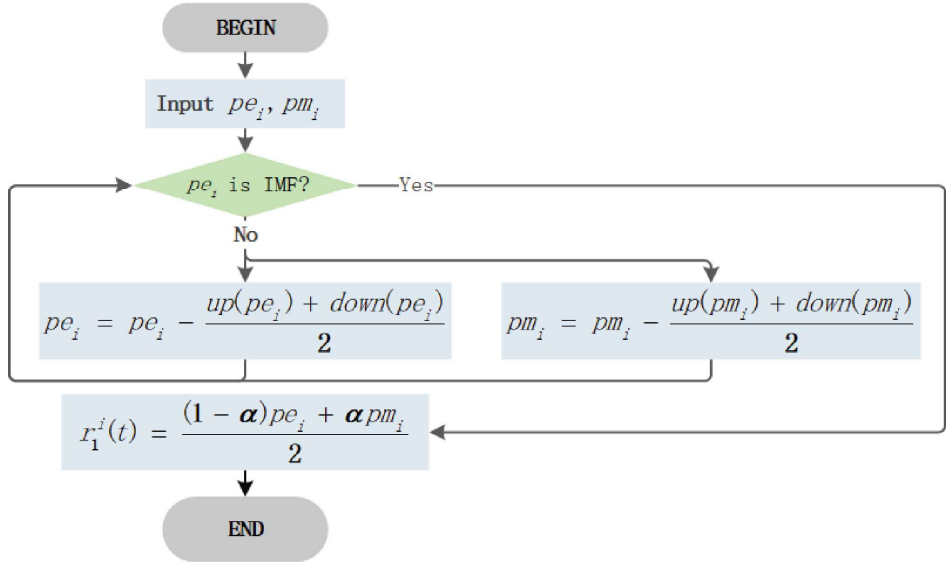

**Fig. 5.** The flowchart of the core function $AM(pe_i, pm_i)$ of ACEEMD.

The core of ACCEMD is $r_1^i(t)$ shown in the Fig. 5, which is referred to as the aliased complete algorithm. It applies cubic interpolation sampling to $pe_i$ and $pm_i$, and the termination condition of the loop in it is that the intermediate signal of $pe_i$ is IMF. The $i$-th first-order IMF component is obtained by taking a weighted average of the intermediate signals of $pe_i$ and $pm_i$, with a default weight $\alpha = 0.5$.

### 2.3 Time-Aware Mechanism

The time-aware mechanism is constructed by linear regression. It is the bias of the distillation module and generates a matrix of the same size as the max pooling result. As part of the distillation module output, it can increase the feature content and minimize information loss of the output features.

Let $W_t$ denote the weight, which is used to multiply the input matrix, $b_t$ denotes the bias matrix, $T$ denotes the matrix of the time-aware mechanism, and $X_{pre}$ is defined as (1). So $T = X_{pre} \times W_t + b_t$. Because the input data of the time-aware mechanism is the same as the data received by probability attention, it can effectively extract features from input data with complete features.

We can express the $i$-th row and $j$-th column of the output $D$ from the distillation module, represented by (2),

$$D_{ij} = \max_{0 \leq m < k, 0 \leq n < k} (f * g)(\bar{\mathcal{A}}(X_{pre}))_{i \times k + m, j \times k + n} + T_{ij} \qquad (2)$$

where, $\bar{\mathcal{A}}$ denotes the probability attention operator, and $k$ denotes the window length of the max pooling. The feature dimension of the distillation module output is halved, so that $k = 2$.

## 3 Experiments

### 3.1 Datasets

We evaluate the proposed method on two real-world datasets, which are NASDAQ100 and SPY500 [24], from US stock markets spanning over ten years. The NASDAQ100 is a stock market index made up of 102 equity stocks of non-financial companies from the NASDAQ. The SPY500 is Standard and Poor's 500, which is a stock market index tracking the stock performance of 500 large companies listed on stock exchanges in the United States.

We selected historical data[1] ranging from Jan-03-2012 to Jan-28-2022 for our experiments. First, we aligned the trading days in the history by removing the lack of data during weekends and public holidays. Then, we split the historical data into training set (Jan-03-2012 to Jun-25-2021), validation set (Jun-28-2021 to Sept-07-2021), and testing set (Sept-07-2021 to Jan-28-2022). In our experiments, we also include mainstream indices (DJIA and NASDAQ) as secondary data when using the two datasets.

### 3.2 Model Setting

In order to avoid the impact of randomly initialized parameters on the prediction results during the training process and obtain stable experimental results we train the model with multiple times. In the experiment of each model, we first train five model results independently using the training set, then we select the model result with the best experimental index performance in the validation set, and finally we use the selected model result to predict the test set.

### 3.3 Evaluation Metrics

**Trend.** We evaluate the performance of forecast trends with two metrics, Accuracy (Acc) and Matthews Correlation Coefficient (MCC) [6] of which the ranges are in $[0, 100]$ and $[-1, 1]$. Note that better performance can be get by higher value of the metrics.

**Return.** Sawhney [24] points out that classification task evaluation metrics can not prove the actual performance of the solution in terms of profit. Therefore, as Sawhney did, we also introduced investment return ratio **(IRR)** [24] and the Sharpe Ratio **(SR)** as metric for solution return. The IRR is defined as $IRR = \sum_{i=1}^{n} R_i + 1$ where $R_i$ denotes the ratio of profit on day $i$ with the range $[-100\%, 100\%]$. The SR is a measure of the return of a portfolio compared to per unit of risk. It is defined as $SR = \frac{E[R_a - R_f]}{std[R_a - R_f]}$ where $R_a$ is the earned return and $R_f$ is risk-free of US[2].

---

[1] https://www.investing.com/.

### 3.4 Competing Methods

We will select the top three models on the authoritative time series prediction leaderboard[3], TimesNet [21], Non-stationary Transformer [23], and DLinear [22], as well as the Informer [20] which using the probability self-attention mechanism as comparison models.

## 4 Result

### 4.1 Trend Evaluation

We use five models to conduct trend prediction experiments on two datasets. The prediction curve of the test set is shown in the Fig. 6.

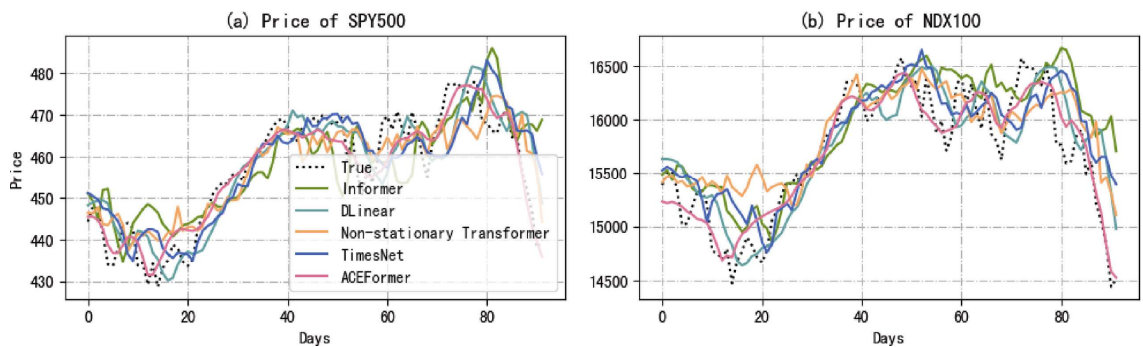

**Fig. 6.** The result of stock trends forecasting by five different models.

In order to quantitatively evaluate the prediction effect of each model, this article uses four indicators for evaluation, and the results are shown in the Table 1. The best results are shown in **bold**.

According to the experimental results in Table 1, it can be clearly seen that among all experimental models, ACEFormer performs the best. In terms of trend evaluation metrics, the ACC and MCC results of ACEFormer on the SPY500 (NASDAQ100) dataset are 69.23% and 0.379 (69.23% and 0.382), respectively. In contrast, only the ACC of the Non-stationary Transformer is slightly larger than 60%, and only the MCC of the Non-stationary Transformer is larger than 0. The ACC intuitively indicates that the fitting degree of the ACEFormer prediction curve is better than other models. At the same time, the MCC proves that ACEFormer can better predict the rise and fall. Based on the ACEFormer prediction results, IRR and SR on the SPY500 (NASDAQ100) dataset are 16.62% and 5.71 (22.31% and 6.43), respectively. In contrast, only the performance of Non-stationary Transformer is better than Benchmark. The IRR shows that ACEFormer can achieve a return of 16.62% (22.31%) on SPY500 (NASDAQ100) within one hundred trading days, and at the same time it can achieve an excess return of 5.71 (6.43) when undertaking a unit of risk. This means that ACEFormer can predict the rise and fall more timely, provide better buying and selling opportunities for trading, obtain greater benefits, and avoid greater losses.

---

[2] https://home.treasury.gov/.
[3] https://github.com/thuml/Time-Series-Library.

**Table 1.** Standard

| Model | SPY500 | | | | NASDAQ100 | | | |
|---|---|---|---|---|---|---|---|---|
| | ACC | MCC | IRR | SR | ACC | MCC | IRR | SR |
| Benchmark | | | −0.97% | −0.27 | | | −5.68% | −0.80 |
| Informer [20] | 43.96% | −0.145 | −8.38% | −2.05 | 45.05% | −0.110 | −10.67% | −1.93 |
| DLinear [22] | 48.35% | −0.041 | −2.65% | −0.90 | 49.45% | −0.021 | −5.76% | −1.22 |
| TimesNet [21] | 48.35% | −0.028 | −6.86% | −1.97 | 45.05% | −0.105 | −8.52% | −1.59 |
| Non-stationary Transformer [23] | 56.04% | 0.116 | 2.13% | 0.48 | 60.44% | 0.195 | −0.86% | −0.23 |
| ACEFormer (Ours) | **69.23%** | **0.379** | **16.62%** | **5.71** | **69.23%** | **0.382** | **22.31%** | **6.43** |

The above statement indicates that in the field of stock prediction, the ACEFormer model performs better than other state-of-the-art models. There are three reasons. First, the ACEEMD algorithm can eliminate as much noise as possible in stock data and reduce the difficulty of predicting long-term trends. Second, the cross-use of multiple attention mechanisms further optimizes feature extraction capabilities. Third, the time-aware mechanism can retain more stock position features and strengthen the temporal coherence of overall features.

### 4.2 ACEEMD Effect

To elaborate on the impact of ACEEMD, we have presented evidence of its effectiveness on stock data of various lengths. Since the unit length of stock data in our solution is 30, we use a 30-day segment of the NASDAQ100 closing price as an example to illustrate the effect of ACEEMD, as shown in Fig. 7. To facilitate the description of ablation experiments, we name ACEEMD without middle points as ECEEMD.

From the endpoints of Fig. 7(a), it is evident that the denoised data obtained by EMD has a significant deviation from the original curve, which is the endpoint effect. On the other hand, the other two denoising algorithms can effectively avoid this issue. Moreover, the curve from day 4 to day 19 is shown in Fig. 7(b). It is observed that the denoised data obtained by ACEEMD can retain the trend of the stock data. In contrast, the other denoised data appear excessively smooth and fail to capture some of the fluctuations presenting in the stock data. Figure 7(c) displays the noise, which is the first-order IMF component, extracted by two different algorithms. It can be observed that the fluctuation trend of the two noises is completely consistent, and at some positions, the fluctuation degree of the noise extracted by ACEEMD is relatively small. This indicates that the positions of noise identified by the two methods are the same. However, ACEEMD is capable of retaining more useful features, and resulting in the preservation of more trends in the stock data.

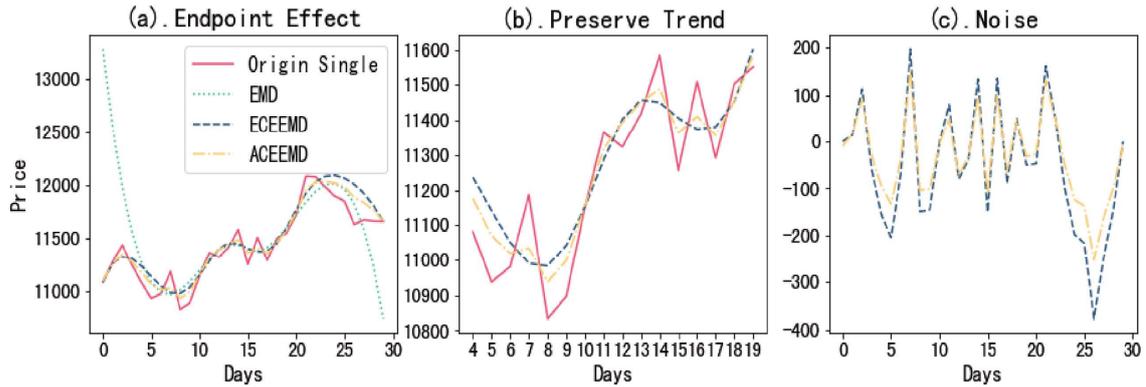

**Fig. 7.** Results of processing stock data using multiple noise reduction algorithms. (a) The effect of the endpoint effect of the EMD algorithm on the noise reduction results. (b) The core function of ACEEMD retaining more stock trends. (c) The noise removed by ECEEMD and ACEEMD respectively.

Thus, we can demonstrate that ACEEMD can not only avoid endpoint effects but also further preserve short-term stock trends.

## 5  Conclusion

In this paper, we address the challenge of predicting nonlinear and highly volatile stock movements and propose a stock trend prediction solution, ACEFormer, that achieves more accurate predictions. In the structure, a denoising algorithm, ACEEMD, is proposed which outperforms existing methods in removing noise from stock data. By using the distillation module and the time-aware mechanism, ACEFormer extracts the key features of denoised stock data and generates more precise predictions. In experimental evaluations, ACEFormer demonstrates improved performance in forecasting stock trends.

**Acknowledgement.** This work is supported in part by the National Nature Science Foundation of China under Grant U2013201 and in part by the Pearl River Talent Plan of Guangdong Province under Grant 2019ZT08X603.